\documentclass[twocolumn,nofootinbib,floatfix,showpacs,preprintnumbers,amsmath,amssymb, prd,10pt]{revtex4}

\usepackage{amsmath}
\usepackage{graphicx}
\usepackage{amssymb}
\usepackage{amsfonts}
\usepackage{latexsym}

\usepackage{graphicx,color}
\usepackage{empheq}
\usepackage{indentfirst}

\def\beq{\begin{eqnarray}}
\def\eeq{\end{eqnarray}}
\def\ln{\,\mbox{ln}\,}

\def\tr{\,\mbox{tr}\,}

\def\Tr{\,\mbox{Tr}\,}

\def\al{\alpha}
\def\be{\beta}

\def\ga{\gamma}
\def\de{\delta}
\def\vp{\varepsilon}
\def\ep{\epsilon}

\def\ka{\kappa}

\def\na{\nabla}

\def\si{\sigma}
\def\om{\omega}

\def\Ga{\Gamma}

\def\La{\Lambda}

\newcommand{\n}[1]{\label{#1}}
\newcommand{\eq}[1]{(\ref{#1})}

\begin{document}

\title{Gauge and parametrization ambiguity in quantum gravity}

\author{Jeferson D. Gon\c{c}alves}
\email{jdgoncalves@fisica.ufjf.br}
\affiliation{
Departamento de F\'{\i}sica, ICE,
Universidade Federal de Juiz de Fora, 36036-330, MG, Brazil}
\author{Tib\'{e}rio de Paula Netto}
\email{tiberiop@fisica.ufjf.br}
\affiliation{Departamento de F\'{\i}sica, ICE,
Universidade Federal de Juiz de Fora, 36036-330, MG, Brazil}
\author{Ilya L. Shapiro}
\email{shapiro@fisica.ufjf.br}
\affiliation{Departamento de F\'{\i}sica, ICE,
Universidade Federal de Juiz de Fora, 36036-330, MG, Brazil}
\affiliation{
Tomsk State Pedagogical University, 634061, Tomsk, Russia}
\affiliation{
National Research Tomsk State University, 634050, Tomsk, Russia}

\date{\today}

\begin{abstract}
\noindent
The gauge and parametrization dependence is discussed in quantum
gravity in an arbitrary dimension $D$. Explicit one-loop calculations
are performed within the most general parametrization of quantum
metric with seven arbitrary parameters. On the other hand, some of
the gauge fixing parameters are fixed to make the calculations relatively
simple. We confirm the general theorem stating that the on shell local
terms in the one-loop effective action are independent of the gauge
and parametrization ambiguity.
\end{abstract}





\maketitle

\section{Introduction}

Loop calculations traditionally play an important role in the
understanding of quantum gravity (QG). The famous pioneer works
in this direction were done by 'tHooft and Veltman \cite{hove},
and Deser and van Nieuwenhuisen \cite{dene} for quantum general
relativity (GR), including the interaction with scalar and vector
quantum fields. It was shown that the one-loop divergences in
pure quantum gravity do vanish on shell, but the interaction with
matter fields always destroys this nice feature. The dependence
on the choice of the gauge fixing conditions was first explored by
Kallosh, Tarasov and Tyutin \cite{KTT}. This complicated calculation
has been performed with a general two-parameter gauge condition.
It was shown that, by means of the gauge-fixing choice, the one-loop
divergences can be reduced to the single topological term which does
not affect the $S$ matrix for gravitons. Of course, this result is
completely consistent with the one \cite{hove} for  the pure
quantum gravity without matter fields or sources.

It is clear that the derivation of divergences, beta functions, and alike
in QG is only the first step, which has not much sense without taking
care of the ambiguities concerning the gauge fixing and, most difficult,
the dependence on the parametrization of the quantum field.

The two-loop calculations in quantum GR \cite{gosa,vanVen} (see also the
recent verification by more advanced methods in \cite{Bern}) confirmed
that even the theory of pure QG is nonrenormalizable. In particular, the
two-loop $S$ matrix can not be done finite in a consistent way. At the
same time, the main
attention was always attracted by the one-loop results, since they
have especially interesting applications. In this respect, one can
mention the asymptotic safety program in QG \cite{MR,RP} and
effective quantum gravity approach \cite{Don94}. In the last case, the
analysis based on the gauge independence of the $S$-matrix elements
proved to be useful \cite{sQED}. After all, we can state that it is
important to know the level of ambiguity for the one-loop divergences
in quantum GR, both the logarithmic and quadratic ones.

The general algorithm to explore the gauge-fixing ambiguities in the
effective action of gauge theories is well-known \cite{VLT82}
(see also \cite{book} for a simplified one-loop version). And since
QG is a particular example of gauge theories, one can easily
establish how the effective action depends on the gauge fixing
condition at the general level and also for the particular gauge fixing
schemes (see, e.g., \cite{lavrov}). At the general level, the issue was
elaborated in the paper of Fradkin and Tseytlin devoted mainly to the
fourth-derivative models of QG \cite{frts82} (see also \cite{avbar86}
and \cite{a} and finally, \cite{Gauss}). In brief, we know that the
one-loop divergences (and also leading divergences at higher loops)
are gauge-fixing independent on the classical mass shell (we call it
simply {\it on shell} in what follows). In principle, the same should
be true for the reparametrization ambiguity. At the same time, it is
sometimes useful to verify the general statements by a direct
calculations, and in the case of QG, this was done in several publications,
at different levels of generality and consistency.  After the pioneer
work \cite{KTT} which explored the gauge-fixing dependence, there
were further publications \cite{Kalm,kalmyk,FirPei} exploring also
the parametrization dependence. In \cite{Kalm,kalmyk}, this was
done by the direct and extremely cumbersome calculation, based
on the heavy use of a computer. The disadvantage of this approach is,
in particular, the fact that this algebra is rather difficult to reproduce.
Contrary to this, in the work of our group \cite{FirPei}, qualitatively
the same result was achieved by a relatively simple handmade
approach, which will be essentially generalized below. In both cases,
it was confirmed that the parametrization dependence vanishes on shell.

Recently, there were some works published which again reconsider the
issue of parametrization and gauge dependence in quantum GR \cite{Falls,RP-2017}. 
The main difference with the previous
papers \cite{KTT,Kalm,kalmyk,FirPei} is that in the publications
\cite{Falls,RP-2017}, the background is not assumed to satisfy the
classical equations of motion. Instead, the background metric has
a special form which is motivated by the arguments of simplicity. In
some cases, it is claimed that there is a gauge-fixing independence
for these special backgrounds. At the same time, the general
statements about ambiguities in gauge theories \cite{VLT82,frts82}
tell us that this independence can be hardly achieved for the most
general choice of parametrization and gauge fixing. Motivated by
these recent works, we extend the previous analysis of  \cite{FirPei}
and consider the most general possible parametrization of a quantum
metric, while the background metric is not constrained. In principle,
our results can be used to reproduce the calculations on any
particular background, being motivated by simplicity, physical
arguments, etc. At the same time, our calculations include a strong
control of correctness, by verifying the general statement of an on shell
universality of the results.

The paper is organized as follows. In Sec.~\ref{s2}, we present a
simple introductory-style analysis of ambiguities of the one-loop
divergences in quantum GR with a cosmological constant. In Sec.~\ref{s3},
one can find the details of the background field method in QG,
including the most general parametrization of a quantum metric and the
most general linear gauge-fixing condition. In Sec.~\ref{s4}, 
the conformal symmetry fixing and the particular form
of the general gauge fixing conditions are described, which makes calculations less
complicated while maintaining almost general choice of parametrization
of quantum variables. The derivation of divergences is reported in
Sec.~\ref{s5}. Furthermore, Sec.~\ref{s6} is about the on shell limit
of the result for the $D$-dimensional quantities which become
logarithmic and quadratic divergences at $D=4$. Finally, in
Sec.~\ref{s7}, we draw our conclusions.

\section{Gauge and parametrization ambiguities in one-loop GR}
\label{s2}

Consider the one-loop effects in GR with a cosmological constant.
For the sake of generality the calculations will be performed in a
generic $D$-dimensional space-time. The metric is supposed to
have a Minkowski signature $(+,-,-,\dots)\,$. Let us note that the use
of heat-kernel methods require the usual analytic continuation to
Euclidean space. We assume this operation without special
explanations. The Einstein-Hilbert action has the form
\beq
\n{EH}
S \,=\, - \frac{1}{\ka^2} \int d^D x \sqrt{-g} \,(R + 2\La)\,,
\eeq
where $\ka^2 \,=\, 16 \pi G$. The equations of motion are
\beq
\vp^{\mu\nu}\,=\,\frac{1}{\sqrt{-g}}\,\frac{\de S}{\de g_{\mu\nu}}
\,=\,R^{\mu\nu}-\frac12\,\big(R+2\La\big)g^{\mu\nu}\,.
\label{vp}
\eeq
Let us use the general statement about gauge-fixing and parametrization
independence on shell for the local part of the effective action.  For the
sake of simplicity, we consider the application of this rule to the
divergences in $D=4$. Then the power-counting arguments tell us that
the divergent part of the one-loop effective action is
\beq
\Ga^{(1)}_{div} &=&\frac{1}{\ep}\,
\int d^4x\sqrt{-g}\{
c_1\,R_{\mu\nu\al\be}^2 + c_2R_{\al\be}^2 + c_3R^2
\nonumber
\\
&+& c_4  {\Box}R + c_5R + c_6  \},
\label{1-loop}
\eeq
where $\,1/\ep\,$ is the divergent coefficient.

According to the Weinberg theorem \cite{Weinberg-QFT}, the ambiguity
in $\Ga^{(1)}_{div}$ leaves this expression local. Then the mentioned
feature of on shell universality tells us that the ambiguity has the form
\beq
\de \Ga^{(1)}_{div}
&=&
\Ga^{(1)}_{div}(\al_i) \,-\, \Ga^{(1)}_{div}(\al^{0}_i)
\nonumber
\\
&=&
\frac{1}{\ep}\,
\int d^4x\sqrt{-g}\,\Big(
b_1 R_{\mu\nu} + b_2 Rg_{\mu\nu} + b_3 g_{\mu\nu} \La
\nonumber
\\
&+&
b_4 g_{\mu\nu}\Box + b_5 \na_\mu \na_\nu\Big)\,\vp^{\mu\nu}\,,
\label{amb}
\eeq
where $\al_i$  represent the full set of arbitrary parameters which
characterize the ambiguity in the choice of gauge-fixing and
parametrization of quantum metric. The special values $\al^{0}_i$
correspond to some special choice of these parameters, e.g., to the
ones which was used in the original paper of 'tHooft and Veltman
\cite{hove}.

The parameters
$b_{1,2,..,5}$ in (\ref{amb}) depend on the choice of $\al_i$, and
the explicit form of the dependence can be known only after the
explicit calculations. However, one can learn a lot about gauge
fixing ambiguity just assuming that the dependence takes place.
In the simplest case without the cosmological constant term,
Eq.~(\ref{amb}) tells us that only the topological Gauss-Bonnet
counterterm can not be set to zero by the special choice of the gauge
fixing condition. This is exactly the result which was first discovered
by direct calculation in the pioneer work \cite{KTT}. The $S$ matrix
corresponds to the on shell limit of effective action, and hence, it is
finite in the theory with $\La=0$.

In the general case of the theory with $\La \neq 0$, the situation is
more complicated. It is
easy to see that the parameter $b_5$ makes no effect on divergences
due to the third Bianchi identity. Therefore, there is a four-parameter
$b_{1,2,3,4}$ ambiguity for the six coefficients $c_{1,2,\dots 6}$.
As a result, only two combinations of these six coefficients can be
expected to be gauge-fixing independent.

Let us elaborate a little bit more on the gauge fixing ambiguity.
Direct calculations show that the parameters of the expression
(\ref{1-loop}) vary according to
\beq
c_1 & \to & c_1\,,
\nonumber
\\
c_2 & \to & c_2+b_1\,,
\nonumber
\\
c_3 & \to & c_3 - \big(b_2+ {\textstyle \frac12}\,b_1 \big)\,,
\nonumber
\\
c_4 & \to & c_4 - b_4\,,
\nonumber
\\
c_5 & \to & c_5 - \big(b_1 + 4 b_2 + b_3 \big)\La\,,
\nonumber
\\
c_6 & \to & c_6 - 4 b_3\La^2\,.
\label{trans}
\eeq
Then, simple linear analysis shows that the two gauge-fixing
invariant quantities are
\beq
c_1
\quad
\mbox{and}
\quad
c_{\rm inv}\,=\,c_6 - 4\La c_5 + 4\La^2 c_2 + 16\La^2 c_3\,.
\label{invs}
\eeq
These two quantities do not modify under the change of the gauge
fixing parameters $\al_i$. It is interesting that the on shell
expressions for the classical action and divergences read
\beq
\n{EH-osh}
S\Big|_{\rm on\,shell}
&=& \frac{6\La}{\ka^2} \int d^4 x \sqrt{-g}\,,
\nonumber
\\
\Ga^{(1)}_{div}\Big|_{\rm on\,shell}
&=& \frac{1}{\ep}\,
\int d^4x\sqrt{-g}\left\{
c_1 R_{\mu\nu\al\be}^2
+ c_{\rm inv}\right\}\,,
\label{1-loop-onshell}
\eeq
and consist only from the gauge-fixing invariant quantities. This fact is
the source of the so-called on shell renormalization group equation, as
noticed in the seminal paper by Fradkin and Tseytlin \cite{frts82}. The
idea can be extended to the Einstein-Cartan model with a cosmological
constant and external spinor current, as was discussed in
\cite{EC-BuSh,Poli}.

The general considerations (see, e.g. \cite{Vilk-Uni}) show that the
expression (\ref{amb}) should also apply to the parametrization
ambiguity, which is in general much more difficult to trace. However,
in this case, the statement is not proved at the same level of safety as
in the case of gauge-fixing dependence \cite{VLT82}, especially in the
situation when two ambiguities are present at the same time. Therefore,
it makes sense to perform explicit calculations and check whether
the property explained above holds in this case. Because of the continuous
interest in the quantum gravity in different dimensions, we perform
this calculation for an arbitrary $D$.

\section{Background-field method for gravity: general setting}
\label{s3}

Our purpose it to perform a derivation of the first two nontrivial
Schwinger-DeWitt coefficients in the most general parametrization
of quantum metric. To this end, using the background field
method, let us consider the following splitting of the metric:
\beq
\n{bgf}
g_{\al\be} &\to& g'_{\al\be}
= e^{2 \ka r\si}
\Big[
g_{\al\be}
+ \ka \big(\ga_1\, \phi_{\al\be}
+ \ga_2\, \phi\, g_{\al\be} \big)
\nonumber
\\
&+&
\ka^2 \big(\ga_3\, \phi_{\al\rho}\phi^\rho_\be
+ \ga_4\, \phi_{\rho\om} \phi^{\rho\om} \,g_{\al\be}
+ \ga_5\,\phi \, \phi_{\al\be}
\nonumber
\\
&+&
\ga_6\, \phi^2 \,g_{\al\be}
\big) \Big],
\eeq
where $g_{\al\be}$ is the background metric and $\phi_{\al\be}$ and
$\si$ are the quantum fields. We also introduce a definition for the trace,
\beq
\phi = \phi^\mu_\mu.
\label{phi}
\eeq
In what  follows, the indexes are lowered and raised with the metric
background $g_{\al\be}$ and its inverse $g^{\al\be}$.

Finally, $\ga_{1,2,\dots,6}$ and $r$ are arbitrary coefficients which
parametrize the choice of the quantum variables. A comment is in order.
As far as the one-loop calculations require only
a bilinear form in the quantum
fields part of the action, it is easy to check that Eq.~(\ref{bgf})
represents the most general possible parametrization of the quantum
metric for the sake of one-loop calculations.

\subsection{Bilinear form in quantum fields}

By using \eq{bgf}, the bilinear form in the quantum fields of action
\eq{EH} reads
\beq
\n{bili}
\hspace{-0.8cm}
&& S^{(2)} =
- \int d^D x \sqrt{-g}\, \Big\{ \phi_{\al\be} \left[
\frac{d_1}{4}\, \de^{\al\be,\mu\nu} \Box
\nonumber \right.
\\
&-& \left.
\frac{d_2}{4}\,  g^{\al\be}g^{\mu\nu} \Box
+ \frac{d_3}{4}\, \left(g^{\mu\nu} \na^\al \na^\be
+ g^{\al\be} \na^\mu \na^\nu \right)
\nonumber \right.
\\
&-& \left. \frac{d_4}{2} \, g^{\be\nu}  \na^\al \na^\mu
- 2 L^{\al\be,\mu \nu} \La
+ \ga_1 ^2 M^{\al\be,\mu\nu}
\right] \phi_{\mu\nu}
\nonumber
\\
&+& \
\phi_{\al\be} \left[
l_0 \, \na^\al \na^\be
+ l_1 \, g^{\al\be} \Box
+ l_2 \, g^{\al\be} \La
\right.
\\
&+& \left. l_3 \,R^{\al\be}
+ l_4 \, g^{\al\be} R \right] \si
+ \si \left[ s_1 \Box +  s_2 \, \La + s_3 R \right] \si \Big\}\,,
\nonumber
\eeq
where the coefficients are as follows:
\beq
d_1
&=& d_4 = \ga_1^2\,,
\nonumber
\\
d_2 &=&
\ga_1^2 + 2\,(D-2)\,\ga_1\ga_2
+ (D-2)(D-1) \,\ga_2^2\,,
\nonumber
\\
d_3  &=&
\ga_1^2 + \,(D-2)\,\ga_1\ga_2\,,
\nonumber
\\
l_0 &=& (D-2) \ga_1  r,
\nonumber
\\
l_1 &=& - (D-2)
\left[ \ga_1 + (D-1)\, \ga_2 \right] r,
\nonumber
\\
l_2 &=&
D \left( \ga_1 + D \ga_2 \right) r,
\nonumber
\\
l_3 &=& - (D-2)  \ga_1 r,
\nonumber
\\
l_4 &=& \frac{(D-2)}{2} \left[ \ga_1 + (D-2) \ga_2 \right] r
\\
&\mbox{and}&
s_1 = - (D-2)(D-1)  r^2,
\quad
s_2 = D^2 r^2 ,
\nonumber
\\
&& s_3 = \frac{(D-2)^2}{2} r^2.
\eeq
In the formula \eq{bili}, the relevant tensor objects are
\beq
\de^{\al\be,\mu\nu} &=& \frac12 \left(
g^{\al\mu} g^{\be\nu} + g^{\al\nu} g^{\be\mu}\right),
\eeq
which is the identity matrix in the space of the symmetric
second-rank fields, and
\beq
M^{\al\be,\mu\nu}
=
\frac{1}{2} R^{\al\mu\be\nu}
- \frac{1+x_1}{4} \de^{\al\be,\mu\nu} R
+ \frac{1+x_2}{2} R^{\al\mu} g^{\be\nu}
\nonumber
\\
- \frac{1+x_3}{4}\left( R^{\al\be} g^{\mu\nu}
 + R^{\mu\nu} g^{\al\be} \right)
+ \frac{1+x_4}{8}
 g^{\al\be}  g^{\mu\nu} R,
\qquad
\eeq
where
\beq
x_1 &=& - \frac{2}{\ga_1^2} \left[\,\ga_3 +
\left( D-2 \right)\ga_4\right]
\,,
\nonumber
\\
x_2 &=& - 2\, \frac{\ga_3}{\ga_1^2}
\,,
\nonumber \\
x_3 &=&  \left( D-4 \right) \frac{\ga_2}{\ga_1}
+ 2\, \frac{\ga_5}{\ga_1^2}
\,,
\nonumber\\
x_4 &=&  2 \left( D-4 \right) \frac{\ga_2}{\ga_1}
+ \left( D-2 \right) \left( D-4 \right) \frac{\ga_2^2}{\ga_1^2}
\nonumber
\\
&+& \frac{4}{\ga_1^2} \left[\ga_5 + (D-2)\ga_6\right]
\eeq
and
\beq
\n{L}
L^{\al\be,\mu\nu} &=& K^{\al\be,\mu\nu}
- \frac12\, (\ga_3 + D \ga_4) \, \de^{\al\be,\mu\nu}
\nonumber
\\
&-&
\frac12\, \left(\ga_5 + D\ga_6\right) g^{\al\be} g^{\mu\nu}\,.
\eeq
Let us explain the condensed notations which were used in these
formulas. In Eq. \eq{L}, there is $K$ tensor
\beq
\n{K}
K^{\al\be,\mu\nu}
&=&
\frac{1}{4} \Big\{\ga_1^2 \de^{\al\be,\mu\nu}
- \frac12 \big[ \ga_1^2 + 2(D-2) \ga_1 \ga_2
\nonumber
\\
&+&
D(D-2) \ga_2^2 \big] g^{\al\be} g^{\mu\nu}
\Big\}.
\eeq
The $K$ tensor is an important object and deserves special
attention. After the introduction of gauge fixing (GF), with a minimal
choice of parameters, the structure \eq{K} will represent the
generalized DeWitt metric in the space of the fields for our model,
see Eq.~\eq{bilipGF}.

Furthermore, in the above formulas and in the following, we used a
special condensed way to write formulas, which enables us to present
the expressions in a relatively compact form. The idea of this
condensed notation is that all the algebraic symmetries are implicit,
including the symmetrization in the couple of indexes
$(\al\be) \leftrightarrow (\mu\nu)$
and inside each couple,
$(\al \leftrightarrow \be)$, $(\mu \leftrightarrow \nu)$.
In order to obtain the complete formulas explicitly, one has to restore
all the symmetries. For example, it is necessary to trade
\beq
R^{\al\mu\be\nu} \to
\frac{1}{4}\,\left(
R^{\al\mu\be\nu} + R^{\al\nu\be\mu}
+ R^{\be\nu\al\mu} + R^{\be\mu\al\nu} \right)
\nonumber
\eeq
and
\beq
R^{\al\mu} g^{\be\nu} 
\to \frac{1}{4}
\big(R^{\al\mu} g^{\be\nu} + R^{\al\nu} g^{\be\mu}
+ R^{\be\nu } g^{\al\mu} + R^{\be\mu } g^{\al\nu} \big)
\nonumber
\eeq
implying that the mentioned symmetries are restored.

\subsection{Gauge fixing action}

Let us introduce the gauge fixing action for the diffeomorphism
invariance in the form
\beq
S_{GF} \,=\, - \frac{1}{\al} \,
\int d^D x \sqrt{-g}\, \chi_\mu \chi^\mu\,,
\eeq
where
\beq
\n{gauge}
\chi_\mu = \na_\rho \, \phi^{\rho}_\mu
- \be_1 \na_\mu \phi - \be_2 \na_\mu \si
\eeq
is the linear background gauge. In the last formulas, $\al\,,\be_1$, and
$\be_2$ are the gauge fixing parameters. The bilinear form of the GF
action is the following:
\beq
\n{biliGF}
&&
S_{GF}^{(2)} =
\int d^D x \sqrt{-g}\,
\Big\{ \phi_{\al\be} \,\Big[\,
\frac{1}{\al} \, g^{\be\nu} \na^\al \na^\mu
\\
&&
- \frac{\be_1}{\al}\,
(g^{\mu\nu} \,\na^\al \na^\be + g^{\al\be} \,\na^\mu \na^\nu)
+ \frac{\be_1^2}{\al}\, g^{\al\be} g^{\mu\nu} \Box \,\Big]
\phi_{\mu\nu}
\nonumber
\\
&&
+ \phi_{\al\be} \,\Big[
\frac{2\be_1 \be_2}{\al}\, g^{\al\be} \Box
-\frac{2 \be_2}{\al}\, \na^\al \na^\be
\Big] \si
+ \frac{\be_2^2}{\al} \, \si \Box \si
\Big\}.
\nonumber
\eeq
By comparing Eqs.~\eq{bili} and \eq{biliGF},
let us note that for the values
\begin{empheq}{align}
\al &= - \frac{2}{\ga_1^2}
\,,
\\
\n{be1}
\be_1 &= \frac{1}{2}
\left[1 + (D-2) \, \frac{\ga_2}{\ga_1}
\right]
\,,
\\
\be_2 &= (D-2)\,\frac{r}{\ga_1}
\end{empheq}
the bilinear operator is minimal. The last means that for these
values of gauge parameters, this operator contains the derivatives
only in the combination \ $\Box = g^{\mu\nu} \na_\mu \na_\nu$.
Then
\beq
\n{bilipGF}
&&
(S + S_{GF})^{(2)} =
- \int d^D x \sqrt{-g}\,
\Big\{ \phi_{\al\be} \big[ K^{\al\be,\mu\nu} \Box
\qquad
\nonumber
\\
&-&
2 L^{\al\be,\mu\nu} \La
+ \ga_1^2 M^{\al\be,\mu\nu}
\big] \phi_{\mu\nu}
\nonumber
\\
&+&
\phi_{\al\be} \big[\tilde{l}_1 g^{\al\be} \Box
+ l_2 \, g^{\al\be} \La
+ l_3 \,R^{\al\be}
+ l_4 \, g^{\al\be} R
\big] \si
\nonumber
\\
&+&
 \si \big[  \tilde{s}_1 \Box
+ s_2 \La + s_3 R \big]  \si
\Big\}
\,,
\eeq
where the new coefficients,
$\tilde{l}_1$ and $\tilde{s}_1$, are
\beq
\tilde{l}_1 &=&  - \frac{(D-2)}{2}\,
(\ga_1 + D \ga_2) \, r,
\nonumber
\\
\tilde{s}_1 &=& - \frac{D(D-2)}{2}\, r^2.
\nonumber
\eeq
It is remarkable and certainly very useful that we could provide the
simplest minimal form of 
a bilinear in a
quantum fields operator for an
arbitrary parametrization of the quantum metric. After that instant,
the calculation becomes pretty much standard, but we shall present
them in full detail, which may be useful for eventual verifications.

\subsection{Trace and traceless decomposition}

It proves useful to separate the field $\phi_{\al\be}$ into trace
(\ref{phi}) and the traceless tensor field,
\beq
\bar{\phi}_{\al\be} \,=\, \phi_{\al\be} - \frac{1}{D} \, g_{\al\be} \phi\,.
\eeq
In the new variables, the bilinear form \eq{bilipGF} becomes
\beq
\n{bili2}
&& (S+S_{GF})^{(2)}
= - \int d^D x \sqrt{-g}
\, \Big\{
\bar{\phi}_{\al\be} \times
\nonumber
\\
&&
\times
\Big[
\frac{\ga_1^2}{4}\, \bar{\de}^{\al\be,\mu\nu} (\Box
- 2 (1+z_1) \La)
+ \ga_1^2 \bar{M}^{\al\be,\mu\nu}
\Big] \bar{\phi}_{\mu\nu}
\nonumber
\\
&&
+ \bar{\phi}_{\al\be} \, [ -2z_2 R^{\al\be}] \, \phi
+ \bar{\phi}_{\al\be}  \, [l_3 R^{\al\be}] \, \si
\nonumber
\\
&&
+ \phi \,[ y_1 \Box  + y_2 \La + y_3 R ]\, \phi
+ \phi  \, [ \tilde{l}_1 \Box + l_2 \La
+ \tilde{l}_3 R ] \,  \si
\nonumber
\\
&&
+ \si \left[  \tilde{s}_1 \Box
+ s_2 \La + s_3 R \right]  \si
\Big\},
\eeq
where the new coefficients $z_{1,2}\,,$
$y_{1,2,3}$, and $\tilde{l}_3$ are
\beq
z_1 &=&
- \frac{2}{\ga_1^2}
\left( \ga_3 + D \ga_4 \right),
\nonumber
\\
z_2  &=& \frac{(D-4) }{4 D}\,
\ga_1 \left(\ga_1 +D \ga_2 \right)
+\frac{\ga_3}{D}+ \frac{\ga_5}{2},
\nonumber
\\
\tilde{l}_3  &=& \frac{(D-2)^2}{2D}
\left(\ga_1 + D \ga_2 \right) r\,,
\nonumber
\\
y_1 &=& - \frac{\left( D-2 \right)}{8D}
\left(\ga_1 + D \ga_2 \right)^2,
\nonumber
\\
y_2  &=&  \frac{\left( D-2 \right)}{4D}
\left(\ga_1 + D \ga_2\right)^2
\nonumber
\\
&+&  \frac{1}{D} \left(\ga_3 + D \ga_4\right)
+ \left(\ga_5 + D \ga_6\right),
\nonumber
\\
y_3  &=& \frac{( D - 2)}{8 D^2} \Big\{
(D-4) \left(\ga_1 + D \ga_2 \right)^2
\nonumber
\\
&+&  4(\ga_3 + D \ga_4)
+ 4 D (\ga_5 + D \ga_6)
\Big\}.
\label{coefs}
\eeq
Also, the projector onto the traceless states is
\beq
\bar{\de}^{\al\be,\mu\nu}
\,=\, \de^{\al\be,\mu\nu} - \frac{1}{D}\, g^{\al\be} g^{\mu\nu}
\eeq
and the last notation is
\beq
\bar{M}^{\al\be,\mu\nu} &=&
\frac{1}{2}\,R^{\al\mu\be\nu}
- \frac{(1+x_1)}{4}\, \bar{\de}^{\al\be,\mu\nu} R
\nonumber
\\
&+&
 \frac{(1+x_2)}{2}\, R^{\al\mu} g^{\be\nu}\,.
\eeq

\section{Conformal gauge fixing}
\label{s4}

In order to remove the remaining degeneracy, let us implement the
conformal gauge fixing in the form
\beq
\si = \be_3 \phi\,,
\eeq
with $\be_3$ being a new free gauge fixing parameter. Let us note that
the conformal gauge fixing does not require Faddeev-Popov ghosts,
because the conformal symmetry transformation has no derivatives
\cite{frts82}. Thus, \eq{bili2} becomes
\beq
\n{bili3}
&&
(S+S_{GF})^{(2)} \,=\, \int d^D x \sqrt{-g}
\, \Big\{
\bar{\phi}_{\al\be} \,
\Big[ \frac{\ga_1^2}{4}\, \bar{\de}^{\al\be,\mu\nu}\Box
\nonumber
\\
&-&
\frac{2\ga_1^2}{4} (1+z_1) \La  \bar{\de}^{\al\be,\mu\nu}
+ \ga_1^2 \bar{M}^{\al\be,\mu\nu} \Big] \bar{\phi}_{\mu\nu}
\nonumber
\\
&+& \bar{\phi}_{\al\be}  \,[ - 2 c R^{\al\be}] \phi
+ \phi [ b_1\Box + 2 b_2 \La
+  b_3 R] \phi
\Big\} \,,
\eeq
where
\beq
c = \frac{D-4}{4D}\, \ga_1
\left(\ga_1 + D \ga_2 \right)
+ \frac{\ga_3}{D} + \frac{\ga_5}{2}
+ \frac{D-2}{2}\, \ga_1 r \be_3
\nonumber
\eeq
and
\beq
b_1 &=& -\frac{D-2}{8 D} \, \big[\ga_1 + D (\ga_2 + 2 r\be_3 )\big]^2,
\nonumber
\\
b_2 &=& \frac{D-2}{8D} \left(\ga_1 + D \ga_2 \right)^2
+ \frac{1}{2D} \left(\ga_3 + D \ga_4 \right)
\nonumber
\\
&+&
\frac12 \left(\ga_5 + D \ga_6 \right)
+ \frac{D}{2} \left(\ga_1 + D \ga_2 \right) r \be_3
+ \frac{D^2}{2} \, r^2 \be_3^2 ,
\nonumber
\\
b_3 &=& (D-2) \left\{
\frac{D-4}{8D^2}  (\ga_1 + D \ga_2 )^2
+ \frac{1}{2 D^2} [ (\ga_3 + D \ga_4 )
\nonumber
\right. \\
&+& \left.
D \left(\ga_5 + D \ga_6 \right) ]
+ \frac{D-2}{2D}\left(\ga_1 + D \ga_2 \right) r \be_3
\nonumber
\right. \\
&+& \left.
 \frac{D-2}{2}\, r^2 \be_3^2
\right\}.
\eeq

\subsection{Bilinear operator in quantum fields}

Now we are in a position to write down the
bilinear in a quantum fields operator in \eq{bili3}
\beq
(S+S_{GF})^{(2)} = - \int d^D x \sqrt{-g}
\left(\begin{array}{cc}
\bar{\phi}_{\al\be} & \phi \end{array} \right)
\hat{H}
\left(\begin{array}{cc}
\bar{\phi}_{\mu\nu}
\\
\phi \end{array} \right),
\eeq
where
\beq
\n{biliH}
\hat{H} &=&
\begin{pmatrix}
\hat{H}_{\bar{\phi}\bar{\phi}} & \hat{H}_{\bar{\phi}\phi}
\\
\hat{H}_{\phi\bar{\phi}}          & \hat{H}_{\phi\phi}
\end{pmatrix},
\qquad
\mbox{and}
\nonumber
\\
\hat{H}_{\bar{\phi}\bar{\phi}}
&=&
\frac{\ga_1^2}{4}\, \bar{\de}^{\al\be,\mu\nu} \big[ \Box - 2 (1+z_1) \La\big]
+ \ga_1^2 \bar{M}^{\al\be,\mu\nu}
\nonumber
\\
\hat{H}_{\bar{\phi}\phi}
&=&
\hat{H}_{\phi\bar{\phi}}
\,=\, - c R^{\al\be}
\nonumber
\\
\hat{H}_{\phi\phi}
&=&
b_1  \Box + 2 b_2\La
+  b_3  R\,.
\eeq

In order to reduce the bilinear form \eq{biliH} into the standard
expression for the minimal operator, $\,\hat{1} \Box + \hat{\Pi}$,
consider a new operator, $\hat{H}' = \hat{C} \cdot \hat{H}$,
where $\hat{C}$ is a $c$-number matrix. Since
\beq
\Tr \ln \hat{H}' \,=\,
\Tr \ln (\hat{C} \cdot \hat{H}) \,=\,
\Tr \ln \hat{C} + \Tr \ln \hat{H},
\eeq
and the contribution of  $\Tr \ln \hat{C}$ does not produce
divergences, i.e., the divergent part satisfies
\beq
\Tr \ln \hat{H}' \big|_{div} \,=\,
\Tr \ln \hat{H} \big|_{div}
\,.
\eeq
By choosing
\beq
\hat{C} \,=\,
\begin{pmatrix}
\frac{4}{\ga_1^2}\, \bar{\de}^{\al\be,\mu\nu}
& 0
\\
0
& \frac{1}{b_1}
\end{pmatrix},
\eeq
we found
\beq
\hat{H}' \,=\, \hat{1} \Box + \hat{\Pi}\,,
\label{SD}
\eeq
where
\beq
\hat{1} \,=\,
\begin{pmatrix}
\bar{\de}^{\al\be,\mu\nu}
& 0
\\
0
& 1
\end{pmatrix}
\eeq
and
\beq
\hspace{-0.2cm}
\hat{\Pi} =
\begin{pmatrix}
4 \bar{M}^{\al\be,\mu\nu} - 2 (1 + z_1) \La\, \bar{\de}^{\al\be,\mu\nu}
& - \dfrac{4c}{\ga_1^2}\, R^{\al\be}
\\
- \dfrac{c}{b_1}\, R^{\mu\nu}
&
\dfrac{2b_2 \La + b_3 R}{b_1}
\end{pmatrix}.
\eeq
The last expression (\ref{SD}) has a standard form, and we can use
known algorithm for the Schwinger-DeWitt technique.

\section{One-loop divergences}
\label{s5}

The one-loop effective action is given by the well-known formula
\beq
\Ga^{(1)} \,=\,
\frac{i}{2} \, \Tr \ln \hat{H} - i \, \Tr \ln \hat{H}_{GH} \,,
\label{trlog}
\eeq
where $\hat{H}$ was defined in previously section and
$\hat{H}_{GH}$ is the Faddeev-Popov ghost operator,
which will be described in the next section.

In $D=2$, the logarithmic divergences in \eq{trlog}
are given by the traces
$\hat{a}_1$ of the coincidence limits of the Schwinger-DeWitt
coefficients  $\hat{a}_1(x,x')$ of the corresponding operators.
In the $D=4$ dimension, $\hat{a}_1$ gives the quadratic divergence,
which is relevant for the applications to asymptotic safety
\cite{MR}, while the traces $\hat{a}_2$ of the coincidence limits of
the Schwinger-DeWitt coefficients $\hat{a}_2(x,x')$ provide
logarithmic operators. For the sake of generality, we will perform
calculations for an arbitrary dimension $D$, which can be also
useful for $2-\ep$ and  $4-\ep$ approaches
and other applications.

\subsection{Derivation of metric contributions}

The next step is to consider the calculation of each term of
Eq.~(\ref{trlog}) separately. According to the Schwinger-DeWitt
technique \cite{BDW-65}
\beq
\n{a2}
\hat{a}_2 & \equiv &
\tr \lim_{x\to x'} \hat   {a}_2 (x,x')
\,=\,
\tr \Big\{ \frac{\hat{1}}{180} \,
\big(R_{\mu\nu\al\be}^2 - R_{\al\be}^2
\nonumber
\\
&+&
\Box R\big)
+ \frac{1}{2} \, \hat{P}^2
+ \frac{1}{6} \,  \Box \hat{P}
+ \frac{1}{12} \, \hat{S}_{\rho\om}^2\Big\},
\eeq
where \ $\hat{P} = \hat{\Pi} + \frac{\hat{1}}{6} R$ \
and, in our case,
\beq
\hat{S}_{\rho\om} \,=\, [\na_\rho,\na_\om] \hat{1}
\,=\,
\begin{pmatrix}
2 g^{\nu\be}\,R^{\,\mu\al}_{\,\,.\,.\,\rho\om}
& 0
\\
0
& 0
\end{pmatrix}.
\eeq
Consequently,
\beq
\n{P}
\hat{P} \,=\,
\begin{pmatrix}
P_{\bar{\phi}\bar{\phi}}^{\al\be,\mu\nu} & P_{\bar{\phi}\phi}^{\al\be}
\\
P_{\phi\bar{\phi}}^{\mu\nu} & P_{\phi\phi}
\end{pmatrix},
\eeq
where
\beq
P_{\bar{\phi}\bar{\phi}}^{\al\be,\mu\nu}
&=&
-\Big[\big( x_1 + {\textstyle \frac56}\big) R
+ 2 (1+z_1) \La \Big]
 \bar{\de}^{\al\be,\mu\nu}
 \nonumber
\\
&+&
2 (1+x_2)\, R^{\mu\al} g^{\nu\be}
+ 2 R^{\al\mu\be\nu},
\nonumber
\\
P_{\bar{\phi}\phi}^{\al\be}
&=&
- 4 \frac{c}{\ga_1^2}\, R^{\al\be},
\nonumber
\\
P_{\phi\bar{\phi}}^{\mu\nu}
&=&
- \frac{c}{b_1} \, R^{\mu\nu},
\nonumber
\\
P_{\phi\phi}
&=&
\frac{2 b_2}{b_1}\,\La
+  \left(\frac{b_3}{b_1} + \frac{1}{6} \right) R\,.
\eeq

In order to evaluate \eq{a2}, let us start from the
$\tr \hat{P}^2$ term. Using \eq{P}, one can write down
\beq
\hat{P}^2 =
\begin{pmatrix}
\hat{P}_{\bar{\phi}\bar{\phi}} \cdot \hat{P}_{\bar{\phi}\bar{\phi}}
+ \hat{P}_{\bar{\phi}\phi}  \cdot \hat{P}_{\phi\bar{\phi}}
&
\cdots
\\
\cdots
& \hat{P}_{\phi\phi} \cdot \hat{P}_{\phi\phi}
 +  \hat{P}_{\phi\bar{\phi}} \cdot \hat{P}_{\bar{\phi}\phi}
\end{pmatrix}.
\nonumber
\\
\n{P2}
\eeq
In this formula, we do not write indexes to clear the notations
and do not show explicitly the irrelevant off diagonal terms. From
(\ref{P2}), it follows
\beq
\n{trP2m}
\tr \hat{P}^2 \,=\, \tr \hat{P}_{\bar{\phi}\bar{\phi}}^2
+ 2 \tr ( \hat{P}_{\bar{\phi}\phi}  \cdot \hat{P}_{\phi\bar{\phi}})
+ \tr \hat{P}_{\phi\phi}^2\,,
\eeq
where the traces are taken in different subspaces of the quantum
metric space.

Introducing the compact notations
\beq
k_1 = \bar{\de}^{\al\be,\mu\nu}
\,, \qquad
k_2 = R^{\mu\al} g^{\nu\be}
\,, \qquad
k_3 = R^{\al\mu\be\nu}
\eeq
we obtain, for the formula \eq{trP2m},
\beq
\n{trP2t}
&& \hspace{-1cm}
\tr \hat{P}^2 =
\Big[
\Big( x_1\, + \frac56 \Big) R
+ 2 (1+z_1) \La \Big]^2 \tr (k_1 \cdot k_1)
\nonumber
\\
&+&
4\, (1+x_2)^2 \, \tr (k_2 \cdot k_2)
\nonumber
\\
&+&
4 \tr (k_3 \cdot k_3)
- 4 \Big[\Big( x_1\, + \frac56 \Big) R
\nonumber
\\
&+&
 2 (1+z_1) \La \Big]
\big[ (1+x_2) \tr (k_1 \cdot k_2) + \tr (k_1 \cdot k_3)\big]
\nonumber
\\
&+&
8(1+x_2) \tr (k_2 \cdot k_3)
+ \frac{8c^2}{b_1\ga_1^2} \,
R_{\al\be} \bar{\de}^{\al\be,\mu\nu} R_{\mu\nu}
\nonumber
\\
&+&
 \Big[ \frac{2b_2}{b_1}\,\La
+ \Big(\frac{b_3}{b_1} + \frac{1}{6} \Big) R \Big]^2.
\eeq

It is not difficult to construct the following multiplication
table for the basic traces
\beq
\tr (k_1 \cdot k_1) & =& \frac{(D-1)(D+2)}{2}\,,
\nonumber
\\
\tr (k_2 \cdot k_2) & =& \frac{(D-2)(D+4)}{4D} \, R_{\al\be}^2
+ \frac{(D^2+4)}{4D^2} \, R^2\,,
\nonumber
\\
\tr (k_3 \cdot k_3) & =& \frac34 \, R_{\mu\nu\al\be}^2
- \frac{2}{D} \, R_{\al\be}^2 + \frac{1}{D^2} \, R^2\,,
\nonumber
\\
\tr (k_1 \cdot k_2) & =& \frac{(D-1)(D+2)}{2D} \, R\,,
\nonumber
\\
\tr (k_1 \cdot k_3) & =&  - \frac{(D+2)}{2D} \, R
\,,
\nonumber
\\
\tr (k_2 \cdot k_3) & =&
-\frac{(D+4)}{2D} \, R_{\al\be}^2 + \frac{1}{D^2} \, R^2\,.
\n{table}
\eeq
We will also need the trace
\beq
R_{\al\be}\, \bar{\de}^{\al\be,\mu\nu}\, R_{\mu\nu}
\,=\,
R_{\al\be}^2 - \frac{1}{D}\,R^2\,.
\eeq

Using the table [Eqs.~(\ref{table})], Eq.~\eq{trP2t} can be evaluated by
using MATHEMATICA \cite{M9}. The result has the form
\beq \n{trP2}
\tr \hat{P}^2 &=& p_1(D) R_{\mu\nu\al\be}^2
+ p_2 (D) R_{\al\be}^2
+ p_4(D) R^2
\nonumber
\\
&+&
 p_5 (D) \La R
+ p_6 (D) \La^2
\,,
\eeq
where
\beq
p_1(D)  &=& 3,
\nonumber
\\
p_2(D)  &=& \frac{D^2-2 D-32}{D}
+ \frac{8 c^2}{b_1 \ga_1^2}
\nonumber
\\
&+&
\frac{D+4}{D} \,
\left[
(D-2)\, x_2^2
+ 2(D-4)\, x_2 \right],
\nonumber
\\
p_4(D)  &=&
\frac{25D^4-95 D^3 + 24D^2+480 D+1152 }{72 D^2}
\nonumber
\\
&+&
\frac{b_3}{3 b_1}
+\frac{b_3^2}{b_1^2}
- \frac{8 c^2}{D b_1 \ga_1^2}
+ \frac{\left(D^2+4\right)}{D^2} \, x_2^2
\nonumber
\\
&+&
(D+2) \Big[
\frac{D-1}{2} \, x_1^2
- \frac{2 (D-1) }{D} \, x_1 x_2
\nonumber
\\
&+&
\frac{5D^2 -17D +24}{6 D}\,x_1
\Big]
\nonumber
\\
&-&
\frac{\left( 5D^3 - D^2 -10D -48  \right) }{3 D^2}\,x_2,
\nonumber
\\
p_5(D)  &=&
\frac{D+2}{D}
\Big[\frac{5D^2- 17D + 24}{3}
\nonumber
\\
&+&
2(D-1) \,(Dx_1 -2x_2) \Big](1+\, z_1)
\nonumber
\\
&+&
\frac{2(b_1 + 6b_3) b_2}{3 b_1^2}\,,
\nonumber
\\
p_6(D)  &=& 2\left(D-1\right)\left(D+2\right) (1+z_1)^2
+ \frac{4b_2^2}{b_1^2}\,.
\eeq

Using formula~\eq{trP2} and the relations
\beq
\tr \hat{1} &=& \frac{D(D+1)}{2}\,,
\nonumber
\\
\tr \hat{S}_{\al\be}^2 &=&  -(D+2) R_{\mu\nu\al\be}^2
\eeq
we arrive at the result for the expression~\eq{a2},
\beq
\n{lima2}
\hat{a}_2 &=& h_1(D) R_{\mu\nu\al\be}^2
+ h_2 (D) R_{\al\be}^2 + h_3 (D) \Box R
\nonumber
\\
&+&
h_4(D) R^2
+ h_5 (D) \La R
+ h_6 (D) \La^2
\,,
\eeq
where
\beq
h_1(D) &=& \frac{\left(D^2-29 D+480\right)}{360}\,,
\nonumber
\\
h_2(D) &=&
-\frac{D^3 - 179D^2 + 360D +5760}{360 D}
\nonumber
\\
&+&
\frac{4c^2}{b_1\ga_1^2}
+\frac{D+4}{D} \Big[\frac{D-2}{2} \,x_2^2 + (D-4) x_2\Big] ,
\nonumber
\\
h_3(D) &=&
-\frac{2D^3-3D^2-5D+20}{30D}
\nonumber
\\
&-&
\frac{(D-1)(D+2)}{12D}\left( D x_1 - 2x_2\right)
+ \frac{b_3}{6b_1},
\nonumber
\\
h_4(D) &=& \frac{25D^4-95 D^3 + 24D^2+480 D+1152 }{144 D^2}
\nonumber
\\
&+&
\frac{b_3}{6 b_1}
+\frac{b_3^2}{2 b_1^2}
- \frac{4 c^2}{D b_1 \ga_1^2}
\nonumber
\\
&+&
(D+2)\Big(
\frac{D-1}{4}\, x_1^2
+ \frac{5D^2 -17D +24}{12 D}\,x_1
\nonumber
\\
&-&
 \frac{D-1}{D} \, x_1 x_2\Big)
+ \frac{\left(D^2+4\right)}{2D^2} \, x_2^2
\nonumber
\\
&-&
\frac{\left( 5D^3 - D^2 -10D -48  \right) }{6 D^2}\,x_2,
\nonumber
\\
h_5(D) &=&
\frac{(b_1 + 6b_3) b_2}{3 b_1^2}
+ \frac{D+2}{D}
\Big[\frac{ 5D^2- 17D + 24}{6}
\nonumber
\\
&+&
(D-1)\,(Dx_1 -2x_2) \Big](1+\, z_1)
\,,
\nonumber
\\
h_6(D)  &=& \left(D-1 \right)\left(D+2\right) (1+z_1)^2
+  \frac{2b_2^2}{b_1^2}\,.
\eeq

A much simpler task is to evaluate
\beq \hat{a}_1
&\equiv&
\tr \lim_{x\to x'} \hat{a}_1 (x,x') \,=\, \tr \hat{P}
\nonumber \\
&=& -\left[
\left( x_1\, + \frac56 \right) R
+ 2 (1+z_1) \La
\right]
\tr (k_1 \cdot k_1)
\nonumber
\\
&+&
2 (1+x_2)\, \tr (k_1 \cdot k_2)
+ 2 \tr (k_1 \cdot k_3)
\nonumber
\\
&+&
2\left(\frac{b_2}{b_1}\right)\,\La
+  \left(\frac{b_3}{b_1} + \frac{1}{6} \right) R
\,.
\eeq
After a small amount of algebra, we find
\beq
\n{a1}
\hat{a}_1 &=&
\Big[\frac{b_3}{b_1} -\frac{5D^3 - 7D^2 -12 D +48}{12 D}
\nonumber
\\
&-&
\frac{(D-1) (D+2) (D x_1 - 2 x_2)}{2 D} \Big] R
\nonumber
\\
&+& \left[ \frac{2b_2}{b_1}
-\left(D-1\right) \left(D+2\right) (1 + z_1)
\right] \La \,.
\eeq

Let us give the expression for divergences in dimensional
regularization for $D \to 4$,
\beq
\frac{i}{2} \, \Tr \ln \hat{H} \big|_{div} \,=\,
- \frac{\mu^{D-4}}{\epsilon}
\int d^4 x \sqrt{-g} \,\, \hat{a}_2
\,,
\eeq
where $\epsilon = (4\pi)^2(D-4)$ and $\mu$ is the dimensional
parameter of renormalization. Consequently,
\beq
&& \frac{i}{2} \, \Tr \ln \hat{H} \big|_{div}
\,=\,
- \frac{\mu^{D-4}}{\epsilon}
\int d^4 x \sqrt{-g} \,
\big\{ h_1 (4)R_{\mu\nu\al\be}^2
\nonumber
\\
&&
+\,
 h_2 (4) R_{\al\be}^2
+ h_3 (4)\Box R
+ h_4 (4)  R^2
\big\},
\eeq
where
\beq
h_1 (4) &=& \frac{19}{18}\,,
\qquad
h_2 (4) = -\frac{55}{18}
+2 x_2^2 +\frac{4 c^2}{b_1 \ga_1^2}\,,
\nonumber
\\
h_3 (4)  &=&
-\frac{2}{3}
+\frac{b_3}{6 b_1}
-\frac{3}{4} \left(2 x_1- x_2\right)\,,
\nonumber
\\
h_4 (4)  &=&
\frac{59}{36}
+\frac{b_3}{6 b_1}
+\frac{b_3^2}{2 b_1^2}
-\frac{c^2}{b_1 \ga_1^2}
\nonumber
\\
&+&
 \frac{9}{2} \left( x_1^2  -  x_1 x_2 + x_1\right)
- \frac{9}{4}\, x_2
+\frac{5 x_2^2}{8}\,,
\nonumber
\\
h_5(4)  &=&
9 + \frac{b_2}{3 b_1}
+ \frac{2 b_2 b_3}{b_1^2}
\nonumber
\\
&+&
9 \left[(2 x_1 - x_2) (1 + z_1) + z_1 \right]\,,
\nonumber
\\
h_6 (4)  &=& 18 (1 + z_1)^2
+ \frac{2 b_2^2}{b_1^2}\,.
\eeq

\subsection{Faddeev-Popov ghost term}

Let us now evaluate the contribution of gauge ghosts.
The Faddeev-Popov ghost operator is defined by
\beq
\n{HGH}
\hat{H}_{GH} &=& \frac{\de \chi^\mu}{\de \phi_{\al\be}}
\, R^{\nu}_{\al\be}
+ \frac{\de \chi^\mu}{\de \si} \, R^\nu
\Bigg|_{\phi_{\al\be} \,\to\, 0,\, \si \,\to\, 0}
\,,
\eeq
where $\chi^\mu$ is the background gauge, defined in
Eq.~\eq{gauge}, and $R^{\nu}_{\al\be}\,$, $\,R^\nu$
are the gauge generators with respect to the quantum fields
$\,\phi_{\al\be}\,$ and $\,\si$, respectively. For the
diffeomorphism symmetry, we have
\beq
\de \phi_{\al\be} \,=\, R^{\mu}_{\al\be} \, \xi_\mu
\,, \qquad \qquad
\de \si \,=\, R^{\mu} \, \xi_\mu
\,,
\eeq
where
\beq
\n{ggph}
R^{\mu}_{\al\be}
&=&
-\frac{1}{\ga_1} \Big(\de_\al^\mu \na_\be + \de_\be^\mu \na_\al
- \frac{2 \ga_2}{\ga_1 + D \ga_2} \, g_{\al\be} \na^\mu \Big)\,,
\nonumber
\\
R^{\mu} &=& \na^\mu \si\,.
\eeq
The details of the derivation of gauge generator \eq{ggph}
can be found in Appendix A. The variational derivatives are
\beq
&&
\frac{\de \chi^\mu}{\de \phi_{\al\be}} \,=\,
\frac12\, (g^{\mu\al} \na^\be + g^{\mu\be} \na^\al)
- \be_1 \,g^{\al\be} \na^\mu
\quad \mbox{and}
\nonumber
\\
&&
\frac{\de \chi^\mu}{\de \si} \,=\, - \be_2 \na^\mu\,.
\eeq
Consequently,
\beq
\n{HGHbe}
&&
\hat{H}_{GH} \,=\,
- \frac{1}{\ga_1} \big(g^{\mu\nu} \Box
+ \tau \na^\mu \na^\nu + R^{\mu\nu} \big),
\nonumber
\\
\mbox{where}
&& \tau = \frac{\ga_1}{\ga_1 + D \ga_2}
\left[1 - 2 \be_1 + (D-2) \frac{\ga_2}{\ga_1} \right].
\eeq
Let us note that the contribution of $\si$ in  \eq{HGHbe}
is irrelevant due to the limit which has to be taken in Eq.~\eq{HGH}.
Now, by using the formula \eq{be1} for the parameter $\be_1$, we get
\beq
\n{gh}
\hat{H}_{GH} \,=\,
- \frac{1}{\ga_1} \left( g^{\mu\nu} \Box + R^{\mu\nu} \right).
\eeq
Indeed, we are lucky enough, that the same choice of gauge fixing
which makes the tensor operator minimal, also makes minimal the
vector operator in the ghost sector.

Using the same logic which was explained for the tensor gravitational
sector, the divergent contribution of the operator \eq{gh} is equivalent to
\beq
\n{gh1}
\hat{H'}_{GH} \,=\,g^{\mu\nu} \Box + R^{\mu\nu}.
\eeq
The expression \eq{gh1} is the minimal vector field operator; hence,
the divergences calculations can be derived, once again by the standard
Schwinger-DeWitt algorithm,
\beq (\hat{a}_1)_{GH}
\,=\, \tr \hat{P}_{GH}\,,
\eeq
\beq
(\hat{a}_2)_{GH}
&=&
\tr \Big[\, \frac{1}{180}\,\hat{1}_{GH} \left(R_{\mu\nu\al\be}^2
- R_{\al\be}^2 + \Box R\right)
\nonumber
\\
&+&
\frac{1}{2} \, \hat{P}^2_{GH}
+  \frac{1}{6} \,  \Box \hat{P}_{GH}
+ \frac{1}{12} \, (\hat{S}_{GH})_{\al\be}^2 \Big],
\eeq
where
\beq
&&
\hat{1}_{GH} \,=\, g^{\mu\nu}\,,
\qquad
\hat{P}_{GH} \,=\, R^{\mu\nu}
+ \frac{1}{6} \, g^{\mu\nu} R \,,
\nonumber
\\
&&
(\hat{S}_{GH})_{\al\be} \,=\, R^{\,\mu\nu}_{\,\,.\,.\,\al\be}\,.
\eeq
Thus,
\beq
&&
\tr \hat{1}_{GH} = D \,,
\qquad
\tr \hat{P}_{GH} =  \frac{(D+6)}{6} \, R\,,
\nonumber
\\
&& \tr \hat{P}_{GH}^2  =  R_{\al\be}^2 + \frac{(D+12)}{36} \, R^2\,,
\nonumber
\\
&&
\tr (\hat{S}_{GH})_{\al\be}^2  \, =\,  -R_{\mu\nu\al\be}^2\,.
\eeq
Finally, we arrive at
\beq
\n{a1GH}
(\hat{a}_1)_{GH} &=& \frac{(D+6)}{6} \, R
\eeq
and
\beq
\n{lima2GH}
(\hat{a}_2)_{GH} &=&
\frac{(D-15)}{180} \, R_{\mu\nu\al\be}^2
- \frac{(D-90)}{180} \, R_{\mu\nu}^2
\nonumber
\\
&+& \frac{(D+5)}{30}\,  \Box R
+ \frac{(D+12)}{72} \, R^2\,.
\eeq
In the limit $D \to 4$, we meet
\beq
- i \Tr \ln \hat{H}_{GH}\big|_{div}
&=&
- \frac{\mu^{D-4}}{\epsilon} \int d^4x \sqrt{-g}
\Big\{
\frac{11 }{90}\,  R_{\mu\nu\al\be}^2
\nonumber
\\
&-& \frac{43 }{45} \, R_{\mu\nu}^2
-\frac{3}{5} \, \Box R
-\frac{4}{9} \, R^2\Big\}.
\eeq
Changing the basis, we arrive at
\beq
&&
- i \Tr \ln \hat{H}_{GH}\big|_{div}
\,=\,  \frac{\mu^{D-4}}{\epsilon} \int d^4x \sqrt{-g}
\Big\{
-\frac{11 }{90}\,  E_4
\nonumber
\\
&+&
\frac{7 }{15} \, R_{\mu\nu}^2
+ \frac{3}{5} \, \Box R +\frac{17}{30} \, R^2 \Big\},
\eeq
where $\,E_4 = R_{\mu\nu\al\be}^2 - 4 R_{\al\be}^2 + R^2\,$
is the 4D Gauss-Bonnet integrand.

\subsection{Divergent part of effective action}

In order to obtain the total value of the $\hat{a}_2$ coefficient, we
need to replace Eq.~\eq{lima2} and Eq.~\eq{lima2GH} into the
general expression (\ref{trlog}). The final answer is similar to
(\ref{1-loop}),
\beq
(\hat{a}_2)_{total} &=&
f_1(D) R_{\mu\nu\al\be}^2
+ f_2 (D) R_{\al\be}^2
+ f_3 (D) \Box R
\nonumber
\\
&+&  f_4(D) R^2
+ f_5 (D) \La R + f_6 (D) \La^2\,,
\eeq
where
\beq
f_1(D) &=& \frac{\left(D^2-33 D+540\right)}{360}\,,
\nonumber
\\
f_2(D) &=&
-\frac{D^3 - 183D^2 + 720D +5760}{360 D}
+\dfrac{4c^2}{b_1\ga_1^2}
\nonumber
\\
&+&
\frac{D+4}{D} \Big[\frac{D-2}{2} \,x_2^2 + (D-4) x_2\Big] ,
\nonumber
\\
f_3(D) &=&
-\frac{2D^3-D^2+5D+20}{30D}
\nonumber
\\
&-&
\frac{(D-1)(D+2)}{12D}\left( D x_1 - 2x_2\right)
+ \frac{b_3}{6b_1}\,,
\nonumber
\\
f_4(D) &=& \frac{25D^4 -99 D^3 - 24D^2+480 D+1152 }{144 D^2}
\nonumber
\\
&+&
\frac{b_3}{6 b_1}
+ \frac{b_3^2}{2 b_1^2}
- \frac{4 c^2}{D b_1 \ga_1^2}
\nonumber
\\
&+&
(D+2) \Big(
\frac{D-1}{4} \, x_1^2
+ \frac{5D^2 -17D +24}{12 D}\,x_1
\nonumber
\\
&-&
\frac{D-1}{D} \, x_1 x_2\Big)
\nonumber
\\
&+&
\frac{\left(D^2+4\right)}{2D^2} \, x_2^2
-\frac{\left( 5D^3 - D^2 -10D -48  \right) }{6 D^2}\,x_2\,,
\nonumber
\\
f_5(D) &=&
\frac{D+2}{D} \Big[\frac{ 5D^2- 17D + 24}{6 }
\nonumber
\\
&+& (D-1)\,(Dx_1 -2x_2) \Big](1+\, z_1)
\nonumber
\\
&+&
\frac{(b_1 + 6b_3) b_2}{3 b_1^2},
\nonumber
\\
f_6(D) &=&
(D-1)(D+2) (1+z_1)^2 + \frac{2b_2^2}{b_1^2}.
\eeq
In the $D\to 4$ limit, we obtain the divergences,
\beq
\n{EAa2}
\Ga^{(1)}_{div} \,=\,
- \frac{\mu^{D-4}}{\epsilon}
\int d^4 x \sqrt{-g} \, (\hat{a}_2)_{total}
\,.
\eeq
One can rewrite \eq{EAa2} in terms of the 4D Gauss-Bonnet term
and the square of the Weyl tensor,
\beq
\n{WeylSq}
C^2 = E_4 +  2 \left(R_{\al\be}^2 - \frac{1}{3} R^2 \right)\,.
\eeq
This can be done by means of the inverse relations
\beq
R_{\mu\nu\al\be}^2 &=& 2 C^2 - E_4 + \frac{1}{3}\,R^2,
\nonumber
\\
R_{\al\be}^2 &=& \frac12\, C^2 - \frac12\, E_4 + \frac{1}{3}\,R^2.
\eeq
After all, the expression for the divergences in an arbitrary
parametrization is
\beq
\n{EACE}
\Ga^{(1)}_{div} &=&
- \frac{\mu^{D-4}}{\epsilon}
\int d^4 x \sqrt{-g} \, \big\{
g_1 C^2 + g_2 E_4 + g_3 \Box R
\nonumber
\\
&+&
g_4 R^2
+ g_5 \La R + g_6 \La^2
\big\}\,,
\eeq
where
\beq
g_1 &=& \frac{7}{20}
+x_2^2 +\frac{2 c^2}{b_1 \ga_1^2},
\nonumber
\\
g_2 &=& \frac{149}{180}
-x_2^2 -\frac{2 c^2}{b_1 \ga_1^2},
\nonumber
\\
g_3 &=&
-\frac{19}{15}
+\frac{b_3}{6 b_1}
-\frac{3}{4} \left(2 x_1- x_2\right),
\nonumber
\\
g_4 &=&
\frac{b_3}{6 b_1}
+\frac{b_3^2}{2 b_1^2}
+\frac{c^2}{3 b_1 \ga_1^2}
+ \frac{9}{2} \left( x_1^2  -  x_1 x_2 + x_1\right)
\nonumber
\\
&-& \frac{9}{4}\, x_2
+\frac{31 }{24}\, x_2^2 + \frac{1}{4},
\nonumber
\\
g_5 &=&
\frac{b_2(b_1 + 6b_3)}{3 b_1^2}
+ 9 \big[(2 x_1 - x_2) (1 + z_1) + z_1 + 1\big],
\nonumber
\\
g_6  &=& 18 (1 + z_1)^2
+ \frac{2 b_2^2}{b_1^2}.
\n{g1-6}
\eeq
For the sake of completeness, the same coefficients are written in
Appendix B in terms of original parameters $\,\ga_{1,...,6}\,$, $\,r\,$,
and $\,\be_3$,  describing  parametrization ambiguity.

Using Eqs.~\eq{a1} and \eq{a1GH}, we can also evaluate the
$\hat{a}_1$ coefficient. The result of this calculation is
\beq
\n{a1t}
(\hat{a}_1)_{total} &=&
\Big[\frac{b_3}{b_1}
-\frac{5D^3 - 3D^2 +12 D +48}{12 D}
\nonumber
\\
&-&
\frac{(D-1) (D+2) (D x_1 - 2 x_2)}{2 D}
\Big] R
\\
&+& \Big[ \frac{2b_2}{b_1} - (D-1) (D+2) (1 + z_1) \Big] \La.
\nonumber
\eeq

\section{Analysis of the results:  known limits and going on shell}
\label{s6}

Let us first consider some special cases of our general answer,
Eq.~\eq{EACE}. First of all, in the limit
\begin{empheq}{align}
x_{1,2}& \to 0
\,,
\qquad
z_1 \to 0
\,,
\qquad
c \to 0
\,,
\nonumber
\\
b_1 & \to -\frac{1}{16}
\,,
\qquad
b_2  \to \frac{1}{16}
\,,
\qquad
b_3  \to 0
\,,
\end{empheq}
one should expect to reproduce the results for GR divergences in
the simplest minimal gauge and simplest parametrization. In fact, we
get in this limit
\beq
\Ga^{(1)}_{div}
&=&
- \frac{\mu^{D-4}}{\epsilon}
\int d^4 x \sqrt{-g} \, \Big\{
\frac{7}{20}\, C^2
+ \frac{149}{180}\, E_4
- \frac{19}{15} \, \Box R
\nonumber
\\
&+&
 \frac{1}{4}\, R^2
+ \frac{26}{3} \La R
+ 20\, \La^2
\Big\}.
\eeq
Using the relation \eq{WeylSq}, this expression becomes
\beq
\Ga^{(1)}_{div}
&=&
- \frac{2\mu^{D-4}}{\epsilon}
\int d^4 x \sqrt{-g} \, \Big\{
\frac{53}{90}\, E_4
+\frac{7}{20}\, R_{\mu\nu}^2
\nonumber
\\
&+&
\frac{1}{120} \, R^2
+ \frac{13}{3} \, \La R +10 \La^2
\Big\},
\eeq
which is the famous result of 'tHooft and Veltman \cite{hove}.
Furthermore, in the limit
\beq
x_{1,2} \to 0 \,,
\qquad
z_1 \to 0\,,
\qquad
c \to \be_3\,,
\nonumber
\eeq
and
\beq
b_1  &\to& -\frac{1}{16} - \be_3  - 4 \be_3^2\,,
\nonumber
\\
b_2 &\to& \frac{1}{16} + 2\be_3  + 8 \be_3^2\,,
\nonumber
\\
b_3 &\to& \frac{1}{2} \be_3 + 2 \be_3^2
\eeq
we checked that the result coincides with the one of Peixoto, Firme
and Shapiro \cite{FirPei}.

\subsection{On shell analysis near $D=4$}

Certainly, the most interesting part is the on shell analysis.
The Einstein equations
\beq
R_{\mu\nu} - \frac{1}{2}\,g_{\mu\nu}
\left(R+ 2 \La \right) \,=\, 0
\eeq
lead to he following relations:
\beq
&& R_{\al\be}^2 \,=\, 4 \La^2, \qquad
R^2 \,=\, 16 \La^2, \qquad
\Box R \,=\, 0,
\nonumber
\\
&&
\La R \,=\, - 4 \La^2
, \qquad
C^2 \,=\, E_4 - \frac{8 \La^2}{3}\,.
\eeq
Using these formulas, the Eq.~\eq{EACE} becomes
\beq
\n{OS1}
\Ga^{(1)}_{div}\Big|_{\rm on\,shell}
 &=&
- \frac{\mu^{D-4}}{\epsilon}
\int d^4 x \sqrt{-g} \, \Big\{
\frac{53}{45}\, E_4
\nonumber
\\
&+&
\Big[18 (2 x_1-x_2-z_1)^2 - \frac{224}{15}
\\
&-&
\frac{2 (b_2-2 b_3) (2 b_1 - 3 b_2+ 6 b_3)}
{3 b_1^2} \Big]  \La^2\Big\}.
\nonumber
\eeq
It is not difficult to see that the second term in the integrand
vanishes, because
\beq
2 x_1-x_2-z_1 = 0\,.
\eeq
For the last term, we have
\beq
&&
b_2-2 b_3 \,=\,
\frac{1}{16} \left(\ga_1 + 4 \ga_2 + 8 r\be_3 \right)^2\,,
\nonumber
\\
&&
(2 b_1 - 3 b_2+ 6 b_3) \,=\,
- \frac{5}{16} \left(\ga_1 + 4 \ga_2 + 8 r\be_3 \right)^2
\nonumber
\\
&&
\mbox{and} \quad
b_1^2 \,=\,
\frac{1}{16^2} \left(\ga_1 + 4 \ga_2 + 8 r\be_3 \right)^4
\,.
\eeq
Therefore,
\beq
- \frac{2 (b_2-2 b_3) (2 b_1 - 3 b_2+ 6 b_3)}
{3 b_1^2} \, \La^2
&=& \frac{10}{3} \, \La^2
\eeq
and the expression \eq{OS1} boils down to
\beq
\hspace{-0.2cm}
\n{OS2}
\Ga^{(1)}_{div}\Big|_{\rm on\,shell}
=
- \frac{\mu^{D-4}}{\epsilon}  \int d^4 x \sqrt{-g}
\Big\{\frac{53}{45}\, E_4 - \frac{58}{5} \,  \La^2 \Big\}.
\eeq
All in all, the one-loop divergences in the on shell limit do not
depend of any parametrization or gauge parameters, exactly as
it should be, see Eq.~(\ref{1-loop-onshell}).

Similarly, for the overall $\hat{a}_1$ coefficient, in the on shell
limit, we have
\beq
\n{a1OS}
(\hat{a}_1)_{total} \Big|_{\rm on\,shell}
&=& \Big[\frac{38}{3} +  18(2 x_1-x_2-z_1)
\nonumber
\\
&+& \frac{2(b_2-2 b_3)}{b_1}\Big] \La\,.
\eeq
As it was explained before, the second term in the {\it r.h.s} of
\eq{a1OS} vanishes. For the third term, one meets
\beq
\frac{2 (b_2-2 b_3) }
{b_1} \La
\,=\, - 2 \La
\eeq
and finally,
\beq
(\hat{a}_1)_{total} \Big|_{\rm on\,shell}
\,=\,  \frac{32}{3} \, \La\,,
\eeq
which does not depend on parametrization or gauge parameters. It is
worth to note that the gauge-fixing independence of the same coefficient
in $D=4$ was established before in Ref.~\cite{Nied}. 

\subsection{$D$-dimensional on shell analysis}

Finally, we can analyze the on shell limit in the Schwinger-DeWitt
coefficients for an arbitrary dimension $D$, where they do not
necessary correspond to a divergent part of the effective action. Taking
the trace of Einstein's equations, we have
\beq
R \,=\, -\frac{2D}{(D-2)} \, \La,
\eeq
and consequently, the field equations can be rewritten as
\beq
R_{\mu\nu} \,=\, -\frac{2g_{\mu\nu}}{(D-2)} \,  \La
\,.
\eeq
Using the above equations, we found for the
$\hat{a}_1$ coefficient, in the on shell limit, that
\beq
(\hat{a}_1)_{total} \Big|_{\rm on\,shell}
\,=\,
-\frac{D \left(D^2-3 D-36\right)}{6 (D-2)} \,\La
\eeq
and
\beq
(\hat{a}_2)_{total} \Big|_{\rm on\,shell}
&=&
\frac{\left(D^2 -33D +540\right)}{360}\, R_{\mu\nu\al\be}^2
\\
&+&
\frac{D \left(5D^3 -17D^2 -354 D-720\right) }{180 (D-2)^2}
\,\La^2
\nonumber
\eeq
for the $\hat{a}_2$ coefficient. We can see that both coefficients
are gauge and parametrization independent in the on shell limit in
general $D$-dimensional space-time. This feature is a clear sign of the
importance of the locality in the gauge-fixing and parametrization
independence of the one-loop effective action. The on shell
universality holds for an arbitrary $D$, independent of whether
the corresponding term is finite or divergent.

\section{Conclusions}
\label{s7}

The universality of beta functions and renormalization group flows
in quantum GR is an important issue, due to the applications to
asymptotic safety and effective quantum gravity approaches.  While
the gauge-fixing dependence is controlled by the on shell conditions,
the parametrization dependence is not completely covered, especially
in the gauge theories. This situation makes interesting the explicit
calculations, but such calculations can become incredibly difficult
in a nonminimal parametrization of gauge fixing.

By employing the ``economic'' approach to the one-loop calculations,
we verified the on shell universality of the first local coefficients of
the Schwinger-DeWitt expansion in an arbitrary dimension $D$. For
the first time, the calculation has been done in the most general
parametrization of a quantum metric, while the gauge-fixing parameters
were partially constrained to provide the minimal form of the tensor
operator of a bilinear form of the total action.

While our calculations were performed only for the first two
coefficients  of the Schwinger-DeWitt technique, the on shell
universality of the result indicated that the parametrization and
gauge-fixing independence of the on shell results is due to the
locality of these terms in the  Schwinger-DeWitt expansion.
Therefore, without explicit calculations, one can ensure that further
coefficients $\hat{a}_k$ with $k \geq 3$, are also on shell universal.

Indeed, the on shell universality property was always regarded as a
useful tool in quantum gravity. As a recent example, one can mention
the gauge-fixing independence of the beta functions in
superrenormalizable models of quantum gravity \cite{highderi}, which
opens the way for interesting applications, such as the possibility to
derive an exact and universal beta function for the Newton constant
\cite{SRQG-beta}. Another example is the recent resolution in
Ref.~\cite{Kame} of the long-standing discrepancy between the
calculations in the phenomenologically interesting tensor-scalar
models, which were done in the Einstein \cite{BKK} and Jordan
frames \cite{spec}.  Our present results indicate that this equivalence
can be extended to the finite part of the effective action, at least to the
local part and to the nonlocal sectors which can be in principle
obtained by the summation of the Schwinger-DeWitt expansions.

\section*{Acknowledgements}

\noindent
J.D.G. is grateful to CAPES for supporting his Ph.D. project.
The work of T.P.N. was supported by PNPD program from CAPES.
I.Sh. gratefully acknowledge partial support from CNPq, FAPEMIG
and ICTP.

\appendix

\section*{Appendix A. On the derivation of the action of ghosts}
\label{apA}

Let us expose some details on the derivation of the generator
\eq{ggph}. The background field splitting of the metric can be
written as
\beq
g'_{\al\be} \,=\, g_{\al\be} + \ka \, h_{\al\be}^{(1)}
+ \ka^2 \, h_{\al\be}^{(2)} + \cdots \,,
\eeq
where
\beq
\n{hab1}
h_{\al\be}^{(1)} &=& \ga_1\, \phi_{\al\be}
+ \ga_2\, \phi\, g_{\al\be}\,,
\\
\n{hab2}
h_{\al\be}^{(2)} &=&
\ga_3\, \phi_{\al\rho}\phi^\rho_\be
+ \ga_4\, \phi_{\rho\om} \phi^{\rho\om} \,g_{\al\be}
\nonumber
\\
&+&
\ga_5\,\phi \, \phi_{\al\be}
+ \ga_6\, \phi^2 \,g_{\al\be}
\eeq
and the dots stand for the $\si$-dependent terms, which we are not
taking into account here. The reason is that, according to
Eq.~\eq{HGH}, the gauge transformation of these terms must be
considered separately. Also, all the terms in \eq{hab2} can be safely
ignored because they are of the second order in the quantum field.
Then, the corresponding part of the gauge generator to \eq{hab2}
must be proportional to $\phi_{\al\be}$ and, consequently, gives no
contribution in the $\phi_{\al\be} \to 0$ limit. Therefore, the use of
Eq.~\eq{hab1} is sufficient for our purposes, because the
$\si$-dependent terms are relevant only starting from the second
loop order. The inverse form of the formula is
\beq
\phi_{\al\be} =
\frac{1}{\ga_1}\Big( \de_{\al\be,}^{\quad\, \mu\nu}
- \frac{\ga_2}{\ga_1+D \ga_2} \, g_{\al\be}\, g^{\mu\nu} \Big)
h^{(1)}_{\mu\nu}\,.
\eeq
Consider the infinitesimal coordinate transformation
\beq
x^{\mu} \rightarrow x'^\mu \,=\, x^\mu + \xi^\mu\,.
\eeq
Then,
\beq
\de h^{(1)}_{\mu\nu} \,=\,
- \left(g_{\mu\rho} \na_\nu
+ g_{\nu\rho} \na_\mu \right) \xi^\rho
\eeq
and we finally get
\beq
\de \phi_{\al\be} &=&
- \frac{1}{\ga_1}
\Big[
g_{\al\rho} \na_\be + g_{\be\rho} \na_\al
\nonumber
\\
&-& \frac{2\ga_2}{\ga_1+D \ga_2} \, g_{\al\be} \na_\rho \Big] \xi^\rho\,,
\eeq
which directly leads to the formula \eq{ggph}.


\section*{Appendix B. The divergences in terms of original parameters}
\label{apB}

Our purpose is to write the expressions (\ref{g1-6}) in terms of the original
parameters of parametrization $\,\ga_{1,2\dots,6},\,r\,$ and gauge fixing
$\be_3$. In order to avoid repetitions in formulas, let us introduce the
notations
\beq
A &=& \ga_3 + 2 \ga_5 + 4 r \be_3  \ga_1\,,
\nonumber
\\
B &=& \ga_1 + 4 \ga_2 + 8 r \be_3\,,
\nonumber
\\
C &=&
8 r \be_3(\ga_1+4 \ga_2 + 4 r \be_3)
\nonumber
\\
&+& (\ga_3 + 4 \ga_4)
+ 4(\ga_5 + 4 \ga_6),
\nonumber
\\
D &=& (\ga_1 + 4 \ga_2)^2
+ 2 \left[(\ga_3 + 4 \ga_4) + 4(\ga_5 + 4 \ga_6)\right]
\nonumber
\\
&+&
32  \be_3 \left[r  (\ga_1 + 4 \ga_2) + 4r^2\be_3 \right],
\nonumber
\\
E &=& \frac{ (\ga_3+4 \ga_4)}{\ga_1^2}\,.
\eeq
Then the coefficients can be cast into the form
\beq
g_1 &=& \frac{7}{20}
+ \frac{4 \ga_3^2}{\ga_1^4} - \frac{2A^2}{ \ga_1^2 B^2}\,,
\nonumber
\\
g_2 &=& \frac{149}{180}
- \frac{4 \ga_3^2}{\ga_1^4}
+ \frac{2A^2}{\ga_1^2 B^2}\,,
\nonumber
\\
g_3
&=&
- \frac{19}{15}
+ \frac{3E}{2}
- \frac{C}{6B^2}\,,
\nonumber
\\
g_4 &=& \frac{1}{4}
-\frac{9E}{2}
+ \frac{31 \ga_3^2 + 216 \ga_4 (\ga_3 + 2 \ga_4)}{ 6 \ga_1^4}
\nonumber
\\
&-&
\frac{A^2}{3\ga_1^2B^2} - \frac{C}{6B^2} + \frac{C^2}{2B^4}\,,
\nonumber
\\
g_5 &=& 9 (1 - 2E )^2
- \Big(\frac{1}{3} - \frac{2C}{B^2}\Big) \,\frac{D}{B^2} \,,
\nonumber
\\
g_6 &=&
18 (1 - 2E )^2 +  \frac{2D^2}{B^4} \,.
\label{g1-6pure}
\eeq


\end{document}